\documentclass[preprint,showpacs,preprintnumbers,amsmath,amssymb,prb]{revtex4}
\usepackage{graphicx}
\usepackage{dcolumn}
\usepackage{bm}

\begin{document}

\title{First-principles study of the effects of gold adsorption on the Al(001) surface properties}

\author{N. Zare Dehnavi}
 \affiliation{Physics Department, Islamic Azad University, Science and Research Campus, Tehran, Iran.}
 
\author{M. Payami}%
 \email[Corresponding author: ]{mpayami@aeoi.org.ir}
\affiliation{ Physics Group, Nuclear Science and Technology Research Institute,\\ Atomic Energy Organization of Iran,  Tehran, Iran.
}%

\begin{abstract}
In this work, we have studied theoretically the effects of gold adsorption on the
Al(001) surface, using {\it ab initio} pseudo-potential method in the framework of the density functional theory.
Having found the hollow sites at the Al(001) surface as the most preferred adsorption sites, we have investigated the effects of the Au adsorption with different coverages ($\Theta$=0.11, 0.25, 0.50, 0.75, 1.00~ML) on the geometry, adsorption energy, surface dipole moment, and the work-function of the Al(001) surface. The results show that, even though the work-function of the Al substrate increases with the Au coverage, the surface dipole moment decreases with the changes in coverage from $\Theta=0.11$~ML to $\Theta=0.25$~ML. We have explained this behavior by analyzing the electronic and ionic charge distributions. Furthermore, by studying the diffusion of Au atoms in to the substrate, we have shown that at room temperature the diffusion rate of Au atoms in to the substrate is negligible but, increasing the temperature to about 200$^\circ$~C the Au atoms significantly diffuse in to the substrate, in agreement with the experiment.
\end{abstract}

\pacs{68.43.Bc, 81.65.Mq, 68.47.Gh}

\maketitle

\section{\label{sec1}Introduction}
 Modification of the Al surface properties by deposition of different kinds of atoms has been studied both experimentally\cite{Hothersall,Porteus,Vanhove,Hutchins,Argile,Egelhoff,Petersen} and theoretically.\cite{Berndt,Ohsaki,Kim1,Kim2} Specifically, the deposition of Au atoms on the Al(001) surface, which is the subject of our present work, has been studied experimentally\cite{Egelhoff} and to our knowledge, there is no report in the literature on theoretical work. The analysis of the experiment conducted by Egelhoff~\cite{Egelhoff} had shown that for sufficiently high coverages (around 10~ML), the Au atoms had been deposited on the Al(001) surface in the form of clusters which is different from the layer-by-layer deposition.

In this work, we have studied the effects of deposition of gold atoms with low-coverages ($\Theta=$ 0.11, 0.25, 0.50, 0.75, 1.00~ML) on the Al(001) surface properties. At these low coverages, the clustering effect would be negligible. For this purpose, we have carried out first-principles calculations within the density functional theory (DFT).\cite{HK,KSH}  Our calculations for the clean Al(001) surface shows outward relaxations for the interlayer spacings, consistent with the other {\it ab initio} calculations\cite{Eguiluz,Bohnen,Fall_98,Zheng,Chis,Borg,Dasilva_05,Sferco} and the latest experiment.\cite{Berndt,Petersen}
In the Au/Al(001) system, we have shown that the work-function,
average binding energy, geometries, and induced surface dipole
moments of the Al(001) surface changes with increasing the
surface density of Au adatoms. Since
the electronegativity of Au atom is higher than that of Al atom
(2.4 for Au and 1.5 for Al, See Ref.~\onlinecite{Joel}), adding Au
atoms on the Al surface, pulls more electrons from the substrate and hence increases the magnitude of the surface dipole moment which, in turn, increases the work-function. However, by
increasing the surface density of Au adatoms, the induced dipole
moment shows a decrease from $\Theta=0.11$~ML to $\Theta=0.25$~ML with a sharp dip at $\Theta$=0.25~ML, a maximum
at $\Theta$=0.50~ML, and a slow decrease up to $\Theta$=1.00~ML. We
have explained this behavior by analyzing the distribution of
space charge of electrons and ions in the slab. The organization of this paper is as follows. In
Sec. \ref{sec2} we explain the detailed calculational methods.
Sec. \ref{sec3} is devoted to the results and discussions and
finally, we conclude this work in Sec. \ref{sec4}.

\section{\label{sec2}Calculation Details}
We have used the Quantum-ESPRESSO code\cite{ESPRESSO} for the electronic structure calculations of the bulk Al, Al(001) surface, and Au/Al(001) system. The calculations are based on the DFT and the self-consistent solution of the Kohn-Sham (KS)\cite{KSH} equations. For the exchange-correlation, we have used the GGA of Perdew {\it et al}~\cite{GGAPBE}; and for the valence electrons of the Al and Au atoms, we have used the scalar-relativistic ultra-soft\cite{Vanderbilt} pseudo-potentials.\cite{pseudo} For the Al pseudo-potential, the electronic configuration $3s^23p^1$
with the same cut-off, $r_c=1.76$ a.u.,
for both orbitals; and for the Au pseudo-potential, the electronic
configuration $6s^16p^{0.5}5d^{9.5}$ with respective cut-offs
$r_c=2.4, 2.6, 2.0$ a.u. were used.

To calculate the bulk properties of fcc Al,
we have tested different k-point samplings and energy cut-offs for the plane wave basis, and have performed the Brillouin-zone integrations using the Methfessel-Paxton smearing\cite{Methfessel} with 0.05~Ry of broadening for appropriate set of ($20\times 20\times 20$) Monkhorst-Pack
grid\cite{Monkhorst} and a 38 Ry
cut-off for the plane wave basis set. Using these parameters, we have solved the self-consistent KS equations for different lattice constant values and obtained the equilibrium lattice constants as those minimizing the total energies. We have considered the atoms in their equilibrium state, when the forces on the atoms were converged to within 1~mRy/a.u.

The Al(001) surface is modeled by a
supercell composed of a 7 atomic layers with a ($1\times 1$) surface unit cell (with the surface-cell size taken from the bulk calculation) and a vacuum
region of 15 empty layers. The kinetic energy cut-off for the plane wave basis and the k-point mesh were chosen as 38 Ry and ($20\times 20\times 1$), respectively.

For the Au/Al(001) system, using Eq.~(\ref{eq1}), we have first found the hollow sites as the preferred adsorption sites, and then a 7-layered Al slab was taken as substrate. The Au atoms were positioned above the preferred adsorption sites on the Al(001) surface, and a vacuum region of 13 empty layers was taken. The surface structures of ($3\times 3$)-Au, ($2\times 2$)-Au, ($2\times 1$)-Au, ($2\times 2$)-3Au, and ($1\times 1$)-Au were chosen for coverages $\Theta$=0.11, 0.25, 0.50, 0.75, and 1.00~ML, respectively. In order to preserve the inversion symmetry, we have positioned the Au atoms on both sides of the slab.
For Au adsorption on the Al(001) surface, the average binding energy per Au atom is defined as
\begin{eqnarray}\label{eq1}
  E_{b}^{\rm Au/Al(001)}(\Theta)=&&-\frac{1}{N_{\rm Au}}[E^{\rm Au/Al(001)}(\Theta)\\ \nonumber
   &&-(N_{\rm Au}E_{\rm Au}+m_1m_2E^{\rm Al(001)})],
\end{eqnarray}
in which $N_{\rm Au}$ is the total number of symmetrically adsorbed Au
atoms on the two sides of the slab ($N_{\rm Au}/2$ on each side) and
$E_{\rm Au}$ is the total energy of an isolated Au atom. $E^{\rm Al(001)}$ is
the total energy of a 7-layered Al supercell having a $(1\times 1)$
surface unit cell with one Al atom at each layer, and
$E^{\rm Au/Al(001)}(\Theta)$ is the total energy of a supercell
composed of a 7-layered Al substrate and Au adatoms with coverage
$\Theta$ and the surface structure of Al(001)-$(m_1\times m_2)$-$(\frac{1}{2}N_{\rm Au})$Au. For example, in $\Theta=0.75$~ML, $E^{\rm Au/Al(001)}(\Theta)$ is the total energy of a supercell composed of a 7-layered Al
substrate with 4 Al atoms on each layer and 3 Au atoms at each
side of the slab. The surface structure in this case is Al(001)-$(2\times 2)$-3Au. The binding energy, Eq.~(\ref{eq1}), is the energy that is gained upon adsorption of a free Au atom on the Al(001) surface. In this definition, a positive number indicates that the adsorption is exothermic with respect to the free Au atom; and a negative number indicates that it is endothermic (i.e., adsorption does not take place). To evaluate the energy of an isolated Au atom, $E_{\rm Au}$, we have solved the self-consistent KS equations for an Au atom in a cubic supercell of 20 a.u. side, a ($7\times 7\times 7$) k-point mesh for the Brillouin-zone integrations, and 38 Ry for the cut-off energy of the plane-wave basis. This choice of k-point mesh was adopted so that the k-point density becomes the same as in the Au bulk calculations. However, the result obtained using the Gamma point alone was the same to within 2 decimals.
On the other hand, choosing 7- and
9-layered Al slabs and adsorption of 1 ML of Au atoms in each case, the difference in the binding energies was
within $10^{-3}$ eV, and the difference in the equilibrium distance of Au atom from the Al surface was less than 0.01~$\AA$.
With these testings, we have chosen a 7-layered Al slab as the substrate in our calculations; a
($20\times 20\times 1$) k-mesh, and 38 Ry cut-off energy for
studying the Au adsorption with ($1\times 1$) surface structure which corresponds to $\Theta$=1~ML. In order
to preserve the same k-point density for different coverages, we have used ($7\times
7\times 1$) mesh for ($3\times 3$)-Au ($\Theta=0.11$~ML), ($10\times
10\times 1$) mesh for ($2\times 2$)-Au ($\Theta=0.25$~ML),
($10\times 20\times 1$) mesh for ($2\times 1$)-Au
($\Theta=0.50$~ML), and ($10\times 10\times 1$) mesh for ($2\times
2$)-3Au ($\Theta=0.75$~ML) surface structures.

In our supercell calculations, to calculate the work-function of a given surface, which is defined as the minimum energy needed to remove an electron from a crystal to a point outside that surface, we find the difference between the electrostatic potential in the middle of the vacuum region and the Fermi energy of the slab. For more accurate results, we have considered sufficiently large vacuum region to prevent the overlap of the wave functions of the neighboring slabs.

In order to analyze the bonding behavior for different coverages, we have considered
the difference electron density\cite{Soon}, $\Delta n({\bf r})$, which
is sensitive to the changes in the charge density, and is defined as
\begin{equation}\label{eq2}
\Delta n({\bf r})=n^{\rm Au/Al(001)}({\bf r})-n^{\rm Al(001)}({\bf
r})-n^{\rm Au}({\bf r}),
\end{equation}
in which $n^{\rm Au/Al(001)}({\bf r})$ is the total electron density
of the substrate-adsorbate system, $n^{\rm Al(001)}({\bf r})$ is the
total electron density of clean Al(001) slab, and $n^{\rm Au}({\bf r})$
is the total electron density of the isolated Au adlayer.

To evaluate the induced surface dipole moment due to Au adsorption, we adopt two different methods. In method 1, we use the Helmholtz equation (in Debye units),\cite{Soon} which is defined as
\begin{equation}\label{eq3}
  \mu=\frac{A\Delta W}{12\pi\Theta},
\end{equation}
in which $A$ is the area in $\AA^2$ per ($1\times 1$) surface unit cell, $\Delta W$ is the work-function change in electron volts, and $\Theta$ is the coverage. In method 2, we resort to the self-consistent charge-redistributions
and directly use the integration
\begin{equation}\label{eq4}
\mu_{\rm ind}=\Delta p_z=\int_{z=0}^{z_{\rm vac}}dz\;z\left[\bar{\rho_2}(z)-\bar{\rho_1}(z)\right],
\end{equation}
in which $\bar{\rho_2}(z)$ is the integral of the self-consistent adsorbate-substrate charge density over the surface unit cell ($m\times n$) divided by $mn$; and $\bar{\rho_1}(z)$ is the same average, but here of the superposition of the self-consistent charge density of the isolated substrate and the self-consistent charge density of the isolated Au adlayer. The limits of the integral, $z=0$ and $z_{\rm vac}$ are, respectively, the position of the central layer of the substrate and the distance of the middle of the vacuum region from the central layer in the adsorbate-substrate system. Here, the position of the central layer of the isolated substrate is identified with the position of the central layer of the adsorbate-substrate system and the position of the isolated Au adlayer is identified with the position of the Au adlayer in the adsorbate-substrate system. Therefore, because of fixed positions, the contributions of the ion charges of the Au adlayer and the central layer (in fact, in our case, the central as well as its first neighbor layers) in Eq.~(\ref{eq4}) cancel out. However, there is no such cancelation in the electronic charge density. In other words, in our 7-layered substrate, only the topmost two layers of the substrate contribute in the ionic redistribution. To apply Eq.~(\ref{eq4}), we use the discretized form
\begin{equation}\label{eq5}
  \mu_{ind}\approx\sum_iz_i\Delta q_i+\sum_j(z^\prime_j-z_j)q,
\end{equation}
in which $z_i$ is the distance of the slice of thickness $\delta z=0.01$~$\AA$ from the slab center
and $\Delta q_i$ is the change in the slice electronic charge. The $z^\prime_j$ and $z_j$ are the ionic positions after and
before the Au adsorption, respectively. Here, we have taken the value of $q=+3e$ for the Al ionic charge.
\section{\label{sec3}Results and Discussion}
In order to study the adsorption of Au atoms on Al(001) surface, we had to consider first the bulk Al, clean Al(001), and the isolated Au atom.

\subsection{\label{subsec-a}Bulk Al and clean Al(001) surface}
 Our calculated lattice constant and bulk modulus for the bulk Al is compared with experimental value and other theoretical results in Table.~\ref{tabm1}.
\begin{table}
\caption{\label{tabm1}Lattice constant, bulk modulus, and cohesive energy of Al.}
\begin{ruledtabular}
\begin{tabular}{lccccccc}
     &present work&LAPW\footnotemark[1]&LAPW\footnotemark[2]&LAPW\footnotemark[3]&PP-PW\footnotemark[4]&PP-PW\footnotemark[5]&Experiment \\ \hline
$a_0$~($\AA$)&4.04 &4.04 &4.10 &4.04 &4.06 &4.05 &4.05$\pm$0.10\footnotemark[6]   \\
$B$~(GPa)&73.8 &75 &73 &78 &75 &79 &72.2\footnotemark[7] \\
$E_{\rm coh}$~(eV/atom)&3.48 &3.65 &- &3.60 &3.52 &3.52 &3.39\footnotemark[7]\\
\end{tabular}
\end{ruledtabular}
\footnotetext[1]{Ref.~\onlinecite{Dasilva_05}}
\footnotetext[2]{Ref.~\onlinecite{Khein}}
\footnotetext[3]{Ref.~\onlinecite{Fuchs1}}
\footnotetext[4]{Ref.~\onlinecite{Favot}}
\footnotetext[5]{Ref.~\onlinecite{Fuchs2}}
\footnotetext[6]{Ref.~\onlinecite{Vanhove}}
\footnotetext[7]{Ref.~\onlinecite{Kittel}}
\end{table}
As is seen, our results are in good agreement with other theoretical works and experiment.

In Al(001) surface calculation, we have obtained the relaxed positions of the first and second outermost layers, and using the value of the bulk spacing, we have calculated the interlayer relaxations,
$\Delta_{ij}=(d_{ij}-d_0)/d_0\times 100\%$, between layers $i$ and
$j$ with respect to the bulk spacing ($d_0$=2.02~$\AA$). Our results show that both interlayer spacings,
$d_{12}$ and $d_{23}$, have outward relaxations consistent with the latest LEED experimental result and other {\it ab initio} calculations. The values are presented in Table~\ref{tab1}. Although our relaxation results for 7-layered slab are smaller than the experimental and the 23-layered LAPW results (Ref.~\onlinecite{Sferco}), but, in contrast to the results of Ref.~\onlinecite{Borg} and Ref.~\onlinecite{Dasilva_05}, the trend is in good agreement with the experiment.
\begin{table}
\caption{\label{tab1}Relaxations of clean Al(001) surface. For comparison, other {\it ab initio} results (only GGA) as well as the experimental result are included. $NL$ indicates the number $N$ of atomic layers used in the slab, pp means pseudo-potential method.}
\begin{ruledtabular}
\begin{tabular}{cccccc}
$\Delta_{ij}(\%)$ &7~L-pp\footnotemark[1] & 7~L-pp\footnotemark[2] & 7~L-LAPW \footnotemark[3] & 23~L-LAPW \footnotemark[4] & Experiment \footnotemark[5]\\ \hline
  $\Delta_{12}$&   +0.91        & +0.5   &  +1.477  & +1.51   & +2.0$\pm$0.8   \\
  $\Delta_{23}$&   +0.26        & -0.3   &  -0.057  & +0.42   & +1.2$\pm$0.7  \\
\end{tabular}
\end{ruledtabular}
\footnotetext[1]{Present work}
\footnotetext[2]{Ref.~\onlinecite{Borg}}
\footnotetext[3]{Ref.~\onlinecite{Dasilva_05}. Calculations up to 17~L has been also considered for which $\Delta_{12}=+1.598$ and $\Delta_{23}=+0.436$}
\footnotetext[4]{Ref.~\onlinecite{Sferco}}
\footnotetext[5]{Ref.~\onlinecite{Petersen}}
\end{table}

\subsection{\label{subsec-b}Proper adsorption site on Al(001) surface}
\begin{figure}
\includegraphics[width=5in]{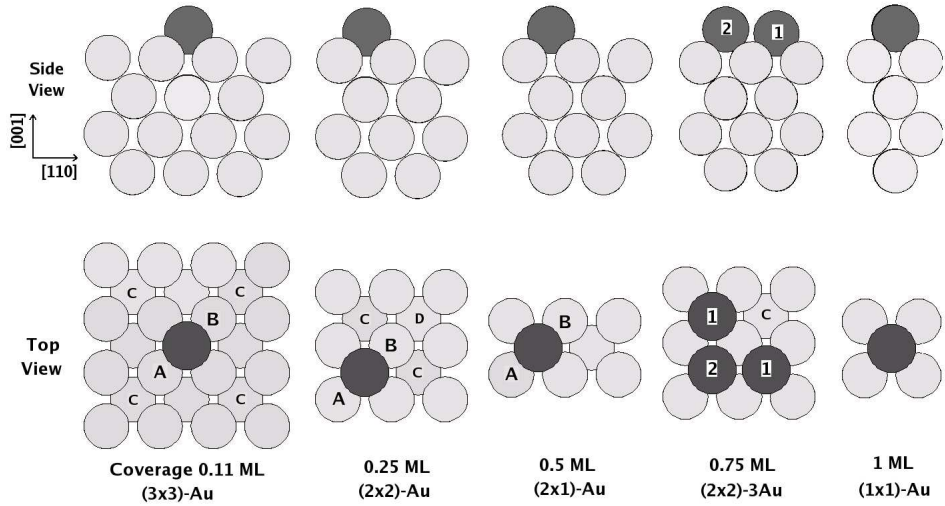}
\caption{\label{fig1}The side and top views of the supercells used in the study of Au atoms on the Al(001) surface.
The black and grey balls represent the Au and Al atoms, respectively. The supercells from left to right correspond to the coverages $\Theta$=0.11, 0.25, 0.50, 0.75, and 1.00~ML, respectively. Only the half of the Al substrate including the central layers are shown. The Al atoms labeled by "C" and "D", and the Au atoms labeled by "1" and "2" show different bucklings. The Al atoms labeled by "A" and "B" are used to define the plane for plots of density profiles as explained in the text. }
\end{figure}

In order to find the proper adsorption site of an Au atom
on the Al(001) surface, we have considered a ($2\times 2$)
surface unit cell for the coverage $\Theta$=0.25~ML, and using Eq.~(\ref{eq1}), we have obtained
the adsorption energies of +2.58 eV/atom, +3.09 eV/atom, and +3.73 eV/atom for
the top, bridge, and hollow sites, respectively. Since all these values are positive, there would be gain of energy in the deposition at each of the three sites. However, since the
resulting values are quite distinct, we choose, confidently, the
hollow sites as the preferred adsorption sites without any need to
repeat the process for larger surface unit cells. We therefore,
consider the adsorption at hollow sites for all coverages. In Fig.~\ref{fig1}, we have shown the side and top views of the surface structures (3$\times$3)-Au, (2$\times$2)-Au, (2$\times$1)-Au, (2$\times$2)-3Au, and (1$\times$1)-Au, which correspond to the coverages of $\Theta$=0.11, 0.25, 0.50, 0.75, and 1.00~ML, respectively. The black balls and grey balls represent Au and Al atoms, respectively.

To study the diffusion of Au atoms into the Al substrate, we have considered a supercell with surface structure of ($2\times 2$) and put an
Au atom at the hollow site under the first layer. Then we allowed the atoms relax
until the force on each atom becomes less than 1 mRy/a.u. The
result showed that the gold atom has pushed the Al atom on its top toward the hollow site above the substrate, i.e., above the "D" atom shown in Fig.~\ref{fig1}, and
has occupied the Al-evacuated position. The total energy of the new equilibrium state (state "$f$") was 0.18~eV higher than the state when the Au atom lies on the substrate (state "$i$"). It is therefore, possible that increasing the temperature enhances this and similar mechanisms which are responsible for diffusion. We have used the nudged elastic band method\cite{Mills} to determine the minimum energy path for the transition from state $i$ to state $f$.
For this purpose, a discretized path consisting of 6 replicas of the system was constructed using linear interpolation between the specified $i$ and $f$ states and let to be optimized iteratively. In Fig.~\ref{fig8} we have shown the variation of the energy along the minimum energy path. The activation energy barrier, $E_a$ for this process which is the difference between the maximum energy along the path and the energy of the initial state, is obtained to be 0.63~eV. We have also used this method to obtain the energy barrier for hopping of Au atom from one hollow site at the surface to the neighboring one and obtained a symmetric curve with $E_a =0.64~eV$, which is of the same order as that of the latter process.
\begin{figure}
\includegraphics[width=4cm,angle=-90]{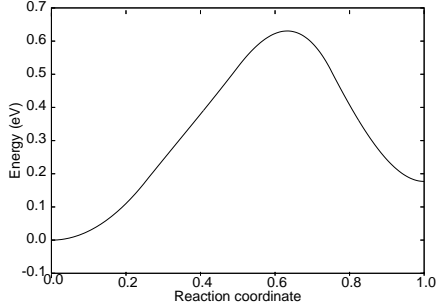}
\caption{\label{fignew}Potential energy along the minimum energy path. The left and right endpoints correspond to states $i$ and $f$, respectively.}
\end{figure}
Assuming that the diffusion can be described within harmonic transition-state theory,\cite{Voter} the transition rate, $1/\tau$, at a temperature $T$ can be written as
\begin{equation}\label{eqnew}
\frac{1}{\tau}=\nu\;e^{-E_a/k_B T},
\end{equation}
in which $E_a$ is the activation energy barrier, and $\nu$ is an effective vibrational frequency which for diffusion on metal surfaces is of order $10^{12}$ s$^{-1}$.\cite{Sorensen} Using the activation energy from Fig.~\ref{fignew}, we obtain the diffusion rate of $\sim 20$ s$^{-1}$ at room temperature, which is negligible; and significant value of $\sim 10^{5}$ s$^{-1}$ at 200$^\circ$~C in good agreement with experiment.\cite{Egelhoff}
\subsection{\label{subsec-c}Geometry of adsorbed Au atoms on Al(001) surface}
We have considered different supercells with their respective surface unit
cells as shown in Fig.~\ref{fig1}, to study the adsorption of the gold atoms  on the Al(001) surface.
In Table~\ref{tab3}, we have presented the relaxations of the first and second layers of the Al(001) substrate for different coverages. In the case of nonzero coverage, since the atoms in the Al(001) layers experience different surroundings, their displacements could be different and therefore, the relaxations are calculated for the mean position of the layers.
\begin{table}
\caption{\label{tab3}Mean relaxations of the Al(001) substrate for different coverages. Here, the average of the atomic positions of a layer in the adsorbate-substrate system is taken as the mean position of the layer. The $\Theta=0.00$~ML represents the clean surface.}
\begin{ruledtabular}
\begin{tabular}{ccccccc}
             &         &       & $\Theta$~(ML)  &          &        &       \\
$\Delta_{ij}(\%)$&  0.00   & 0.11  & 0.25           &  0.50    & 0.75   & 1.00  \\ \hline
$\Delta_{12}$& +0.91   & +0.68 & +0.25 & -0.21   & -0.94  & -1.21   \\
$\Delta_{23}$& +0.26   & +0.15 & -0.43 & +0.59   & -0.58  & +0.01   \\
\end{tabular}
\end{ruledtabular}
\end{table}
As is seen from Table~\ref{tab3}, for $\Theta$=0.11~ML, the relaxations are still as expansions while, in $\Theta$=0.25~ML, the first spacing is expanded and the second is contracted. In $\Theta$=0.5~ML, the first is contracted while the second expanded; in $\Theta$=0.75~ML, both are contracted; and finally, in $\Theta$=1.00~ML, the first spacing is contracted, whereas the second one is expanded. To go further from the mean behaviors, we mention that the atoms of the first layer of Al(001) lie in a plane for all coverages except for $\Theta$=0.11~ML, in which they show a small bucklings ($<<0.01$~$\AA$); and the second layer shows different behaviors for different coverages.
For example, in $\Theta$=1.00~ML, all the atoms in the second layer lie in a plane; in $\Theta$=0.5~ML, the bucklings are vanishingly small ($<<0.01$~$\AA$); in
 $\Theta$=0.11~ML, the atoms labeled by "C" lie 0.01~$\AA$ above the plane and the atoms just under the Au atoms lie 0.02~$\AA$ below the plane while, the displacement of the others are negligible. In $\Theta$=0.25~ML, the "C" atoms lie 0.02~$\AA$ above the plane, the "D" atoms lie 0.01~$\AA$ below the plane, and the atoms just under the Au atoms lie 0.03~$\AA$ below the plane. Finally, in $\Theta$=0.75~ML, the Al atoms under the "1"-labeled Au atoms, Au(1), lie 0.03~$\AA$ below, the Al atoms under the "2"-labeled Au atoms, Au(2), lie 0.03~$\AA$ above, and the "C" atoms lie 0.04~$\AA$ above the second Al plane.
\begin{table}
\caption{\label{tab4}Mean distance of the Au plane from the Al(001) surface for different coverages. The remaining is the same as in Table~\ref{tab3}.}
\begin{ruledtabular}
\begin{tabular}{ccccccc}
             &         &       & $\Theta$~(ML)  &          &        &       \\
                    &  0.00   & 0.11  & 0.25    &  0.50    & 0.75   & 1.00  \\ \hline
$d_{\rm Au-Al}(\AA)$&         & 1.55  & 1.54    & 1.67     & 1.73   & 1.80  \\
\end{tabular}
\end{ruledtabular}
\end{table}
It is seen from Table~\ref{tab4} that the distance of the Au adlayer from the Al substrate almost increases with the coverage. In coverage $\Theta$=0.75~ML, the Au atoms are grouped in two classes. The Au(1) atoms lie 0.06~$\AA$ below the mean Au adlayer plane while, the Au(2) atoms lie 0.12~$\AA$ above that plane.


\subsection{\label{subsec-d}Au adsorption energies for Al(001) surface}
Using Eq.~(\ref{eq1}), we have calculated the adsorption energy of gold atoms on the Al(001) surface for different surface densities (0.11~ML$\le\Theta\le$1.00~ML). The result is summarized in Table~\ref{tab6}. As is seen, the adsorption energy
increases with coverage up to $\Theta=0.5$~ML; and from $\Theta$=0.5~ML to $\Theta$=1.00~ML it has a decreasing behavior. However, the differences are so tiny (at most 0.06 eV) that practically they do not change with the coverage.
\begin{table}
\caption{\label{tab6}Adsorption energies of Au atoms for different coverages.}
\begin{ruledtabular}
\begin{tabular}{cccccc}
     & & & $\Theta$~(ML) &  & \\
    &0.11&0.25&0.50&0.75&1.00   \\ \hline
$E_b~(eV)$& 3.72 &3.73 &3.78 &3.77 &3.76  \\
\end{tabular}
\end{ruledtabular}
\end{table}
\subsection{\label{subsec-e}Work functions and surface dipole moments}
Using the self-consistent electrostatic potentials and the Fermi energies, we have calculated the work functions, $W$; and thereby, using the Helmholtz relation [Eq.~(\ref{eq3})], we have calculated the values of the induced surface dipole moments, $\mu$, of the Au/Al(001) surface with coverages $\Theta$=0.11, 0.25, 0.50, 0.75, 1.00~ML (method 1).
The results are summarized in Table~\ref{tab7}.
\begin{table}
\caption{\label{tab7}Work-functions and change in work-functions, in electron-volts; induced surface dipole moments, in Debye units, of Au/Al(001) surfaces with different coverages, $\Theta$=0.00, 0.11, 0.25, 0.50, 0.75, 1.00~ML. The $\Theta$=0.00~ML refers to the clean surface.}
\begin{ruledtabular}
\begin{tabular}{lcccccc}
                &    &    &$\Theta$~(ML)& & &\\
                &0.00&0.11&0.25&0.50&0.75&1.00\\ \hline
$W$         &4.24 &4.30 &4.32 &4.56 &4.70 &4.81  \\
$\Delta W$  & 0.00&0.06 &0.08 &0.32 &0.46 &0.57  \\
$\mu$\footnotemark[1]& &0.13 &0.07 &0.14 &0.13 &0.12  \\
$\mu_{\rm el}$\footnotemark[2]& &-1.03&-1.04&-0.13&-0.60&-0.63  \\
$\mu_{\rm ion}$\footnotemark[3]& &1.15 &1.10 &0.26 &0.72 &0.75  \\
$\mu_{\rm tot}$\footnotemark[4]& &0.12 &0.06 &0.13 &0.12 &0.12  \\
\end{tabular}
\end{ruledtabular}
\footnotetext[1]{Method 1, calculated using Helmholtz equation [Eq.~(\ref{eq3})]}
\footnotetext[2]{The electronic contribution, the first sum in Eq.~(\ref{eq5})}
\footnotetext[3]{The ionic contribution, the second sum in Eq.~(\ref{eq5})}
\footnotetext[4]{The sum of the electronic and ionic parts (method 2)}
\end{table}

As is seen in Table~\ref{tab7}, from $\Theta$=0.11~ML to $\Theta$=0.25~ML, the change in the work function is quite small (+0.02 eV), while, the work-function increases significantly from $\Theta$=0.25~ML to $\Theta$=1.00~ML. On the other hand, the induced dipole moment shows a decreasing behavior from $\Theta=$0.11~ML to $\Theta=$0.25~ML with a sharp dip at $\Theta$=0.25~ML, an increasing behavior from $\Theta$=0.25~ML to $\Theta$=0.50~ML, and a slow decreasing behavior from $\Theta$=0.50~ML to $\Theta$=1.00~ML. Such this "strange" behavior had been also encountered in the earlier work for adsorption of oxygen on Cu(111) by Soon {\it et al}.\cite{Soon}

\begin{figure}
\includegraphics[width=7cm]{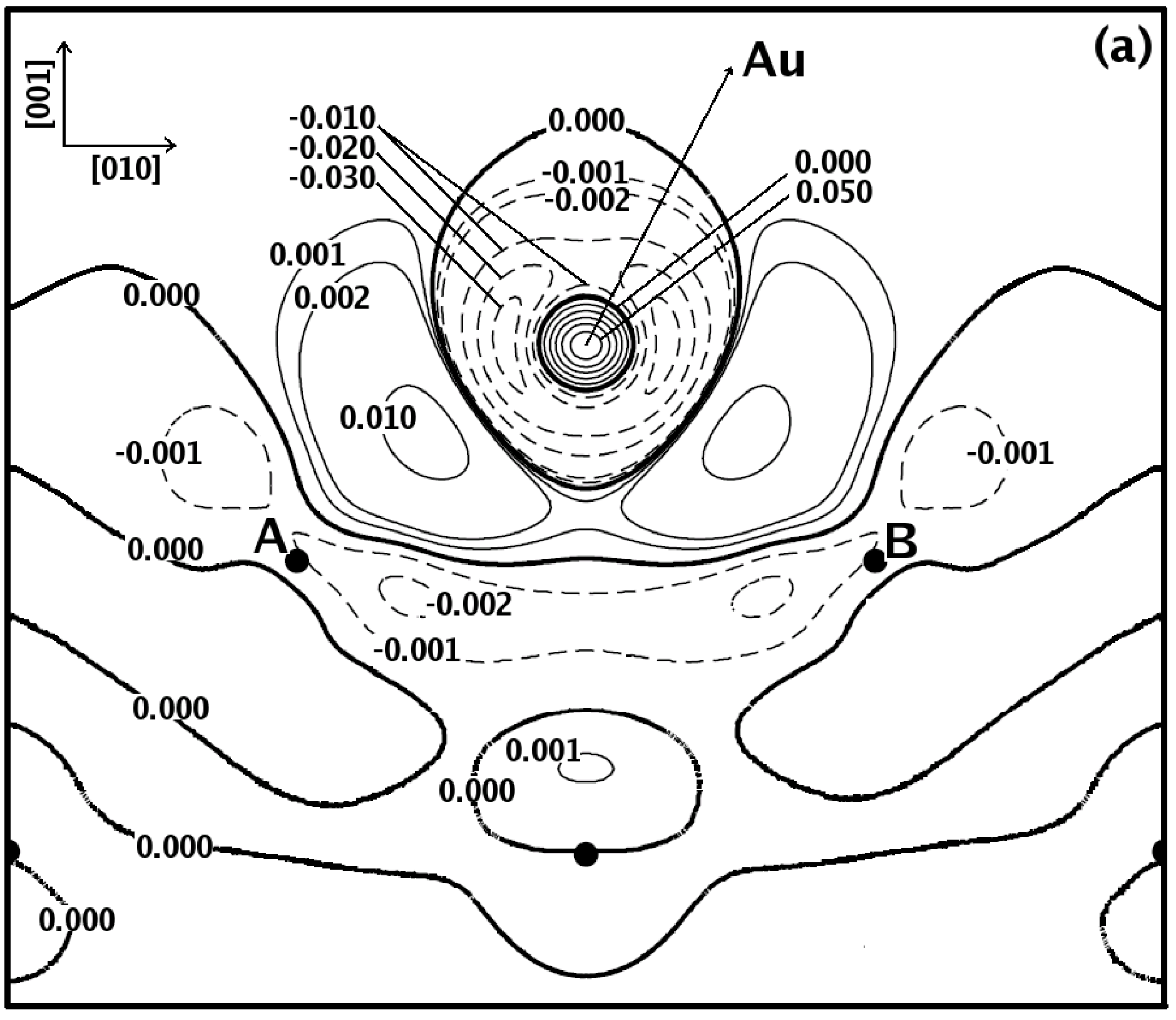}
\includegraphics[width=7cm]{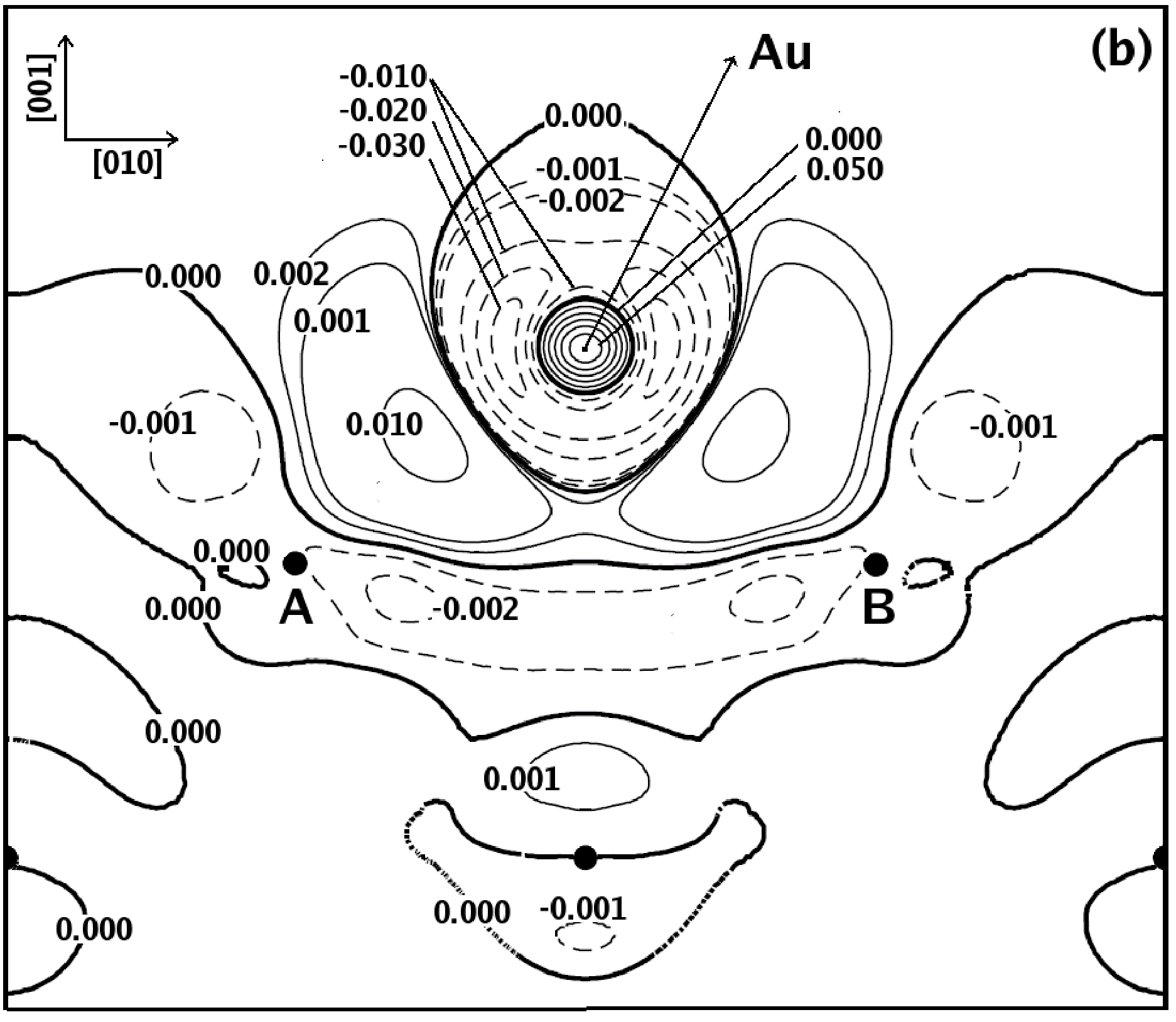}
\includegraphics[width=7cm]{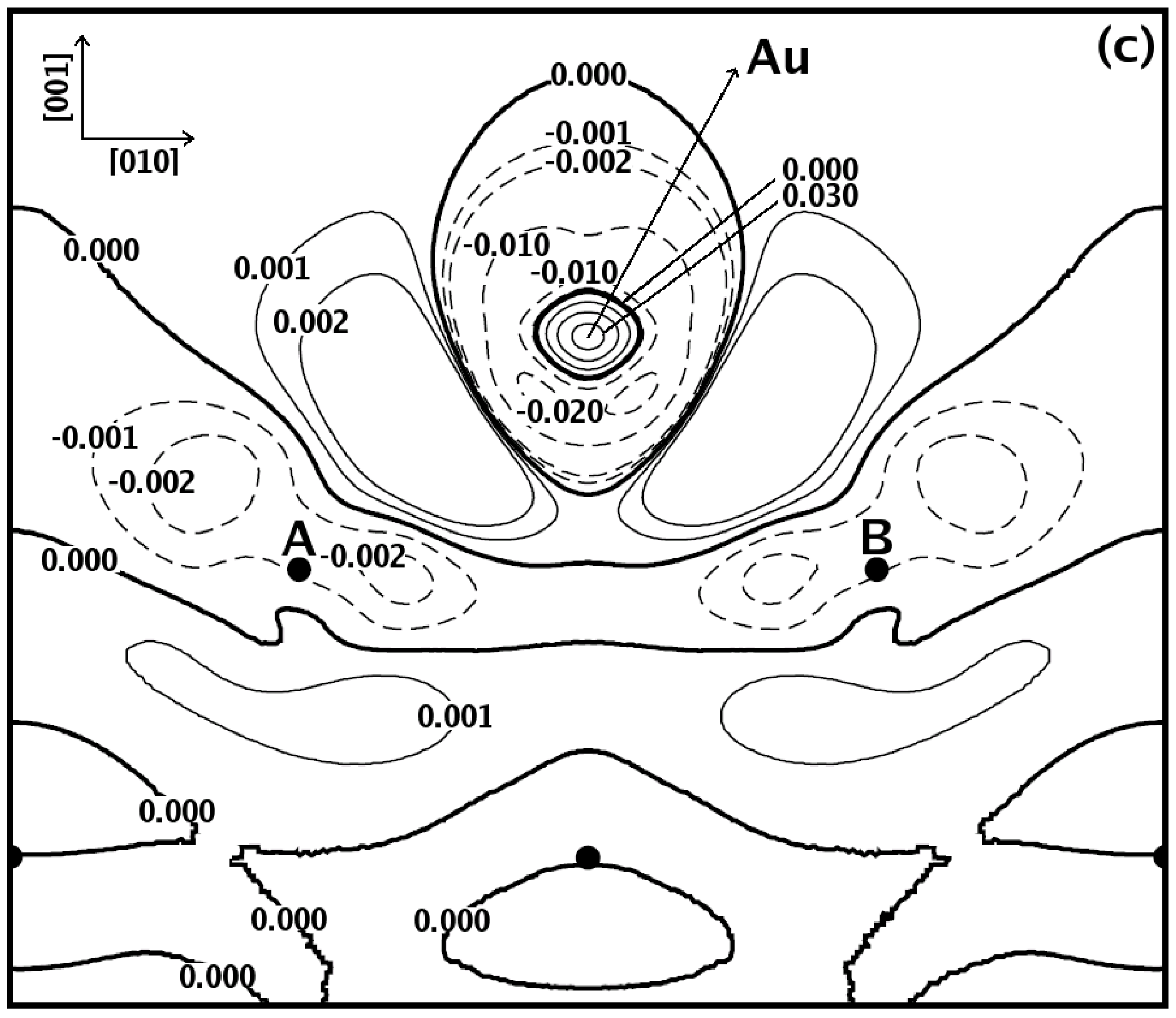}
\caption{\label{fig4}Contour plots of the difference electron densities in the (100) plane
normal to the Al(001) surface which passes through the Au atom and
its two nearest Al(A) and Al(B) atoms (see Fig.~\ref{fig1}). (a), (b), and (c) correspond to
$\Theta$=0.11, 0.25, and 0.50~ML, respectively. Contour levels in (a) and (b) go from -0.03 to 0.05 $e$/(a.u.)$^3$ in steps of 0.01 $e$/(a.u.)$^3$; and in (c), from -0.02 to 0.03 $e$/(a.u.)$^3$ in steps of 0.01 $e$/(a.u.)$^3$, while those between -0.002 and 0.002 $e$/(a.u.)$^3$ were taken in steps of 0.001 $e$/(a.u.)$^3$. Solid (broken) lines indicate positive (negative) values. The thicker solid lines indicate the zero contour levels. The solid circles represent Al atoms.}
\end{figure}

\begin{figure}
\includegraphics[width=4cm,angle=-90]{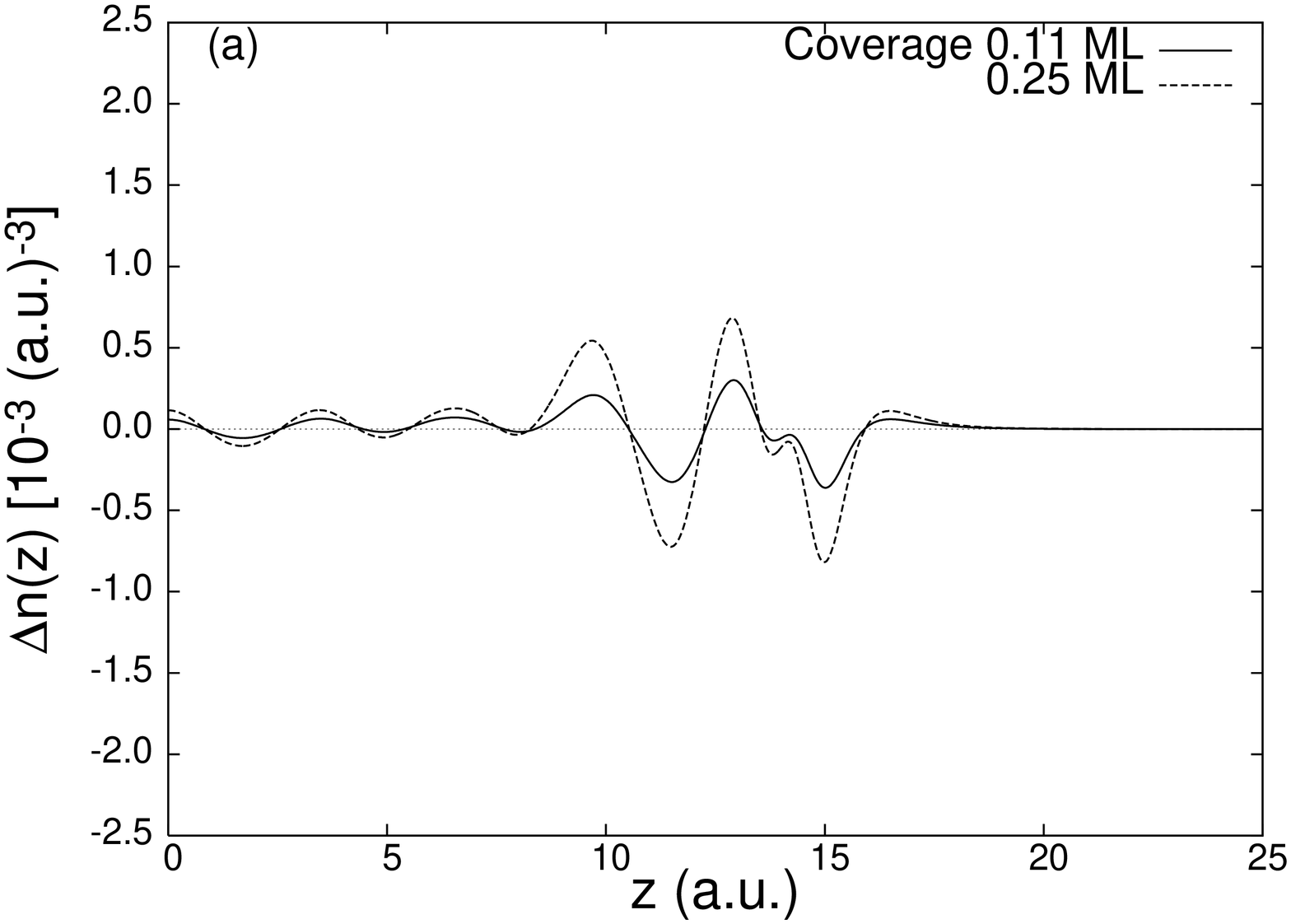}
\includegraphics[width=4cm,angle=-90]{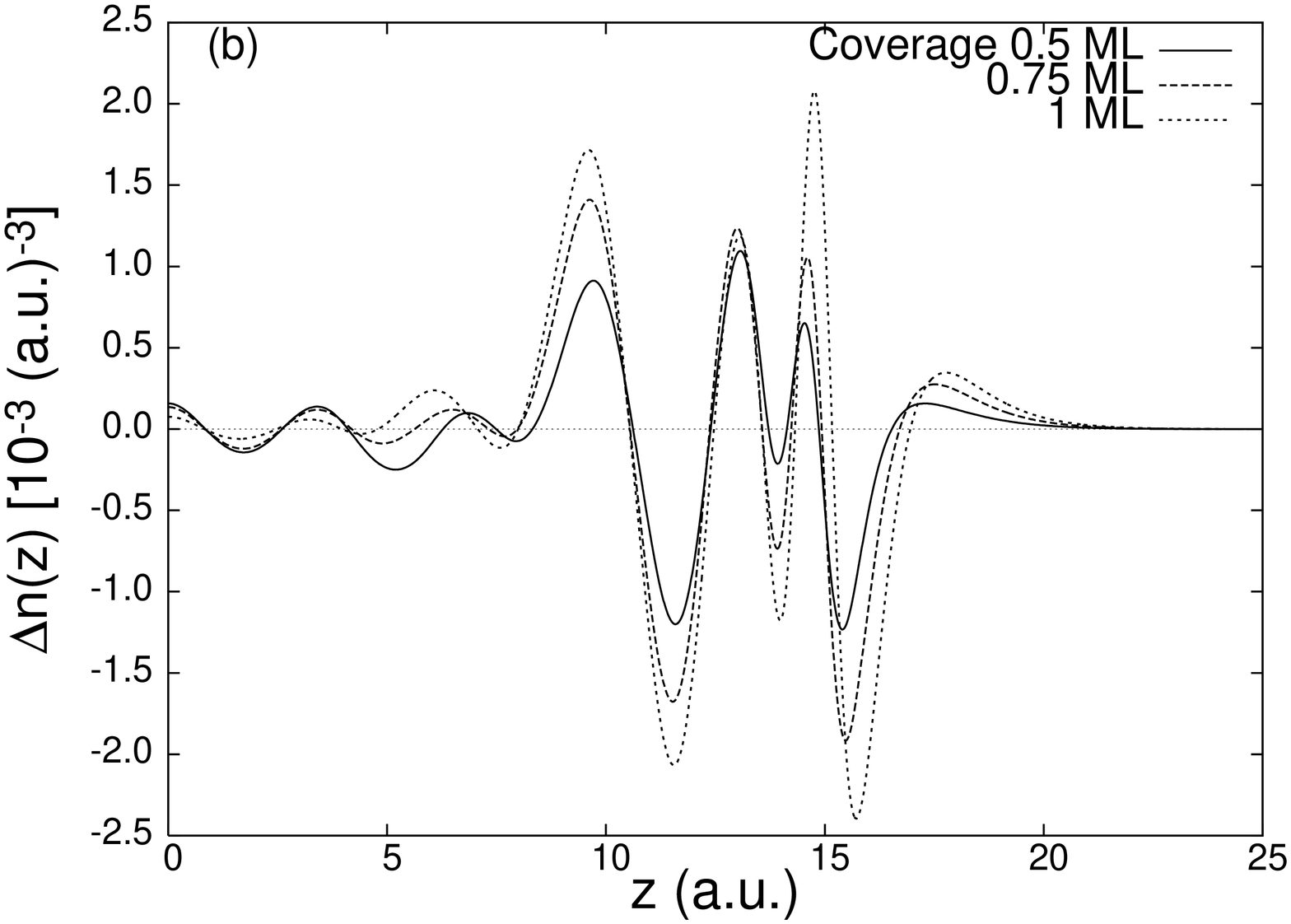}
\caption{\label{fig5}Difference electron densities averaged over the surface unit cell; (a)- for $\Theta$=0.11~ML and $\Theta$=0.25~ML, and (b)- for $\Theta$=0.5, 0.75, 1.00~ML. The positive values imply an increase in the electronic density.}
\end{figure}

To proceed in explaining the sharp dip, we have plotted, in Figs.~\ref{fig4}(a)-(c), the profile of the difference electron charge densities in the plane (100) normal to the Al(001) surface, which passes through the adsorbed Au, substrate Al(A), and Al(B) atoms.
The reason for choosing this plane is that the
difference electronic charge density was higher than that in other planes. As is seen, some features are common in all the three cases. First, there exist a region with decrease and a region with increase of charge around the Au atom; second, the electronic charges partly reside in the region between the Au atom and the Al substrate; and third, there is a decrease of charge around the Al atoms of the first layer. On the other hand, the detailed form of the redistributions in different planes normal to the Al(001) plane differ from each other. However, since only the $z$-component of the dipole moments determines the changes in the work-function, we should take an average over the plane parallel to the surface. In Figs.~\ref{fig5}(a)-(b), we have plotted the averaged difference electron densities as functions of $z$, the normal distance from the central layer of the slab, for different coverage structures. To be clear, we have plotted the $\Theta$=0.11~ML and $\Theta$=0.25~ML cases in Fig.~\ref{fig5}(a), and the remaining cases in Fig.~\ref{fig5}(b).
Inspecting the positions of the adsorbed Au atoms (the average position in the case of $\Theta$=0.75~ML) in the adsorbate-substrate systems reveals that the local maxima to the left of the outermost minima in Figs.~\ref{fig5}(a)-(b) are located at the positions of adsorbed Au atoms. In Fig.~\ref{fig5}(a), the difference densities at the Au atoms are negative (i.e., the Au atom in the adsorbate-substrate system has a less electron density compared to the Au atom in the isolated adlayer) while, in Fig.~\ref{fig5}(b), they are positive. On the other hand, in the case of $\Theta$=0.25~ML, the value of the difference density at the position of the Au atom is minimum compared to the lower and higher coverage structures.
Using the $z$-dependent difference electron densities, and multiplying by the slice thickness ($\delta z=0.01$~$\AA$) and the slice position, $z_i$, we have obtained the electronic contributions, $\mu_{\rm el}$, of the induced surface dipole moments [first term in Eq.~(\ref{eq5})]. The electronic and the ionic contributions as well as the sum of them are presented in Table~\ref{tab7}. It is clear from Table~\ref{tab7} that the electronic contribution has a sharp maximum at $\Theta$=0.50~ML which can not be related to the sharp dip in the induced surface dipole moment. However, adding the ionic part yields the correct values of the induced dipole moments, $\mu_{\rm tot}$, which we had called as method 2.
The results show an excellent agreement with those obtained in method 1 (using Helmholtz relation).
\subsection{\label{subsec-f}Local density of states}
\begin{figure}
\includegraphics[width=4cm,angle=-90]{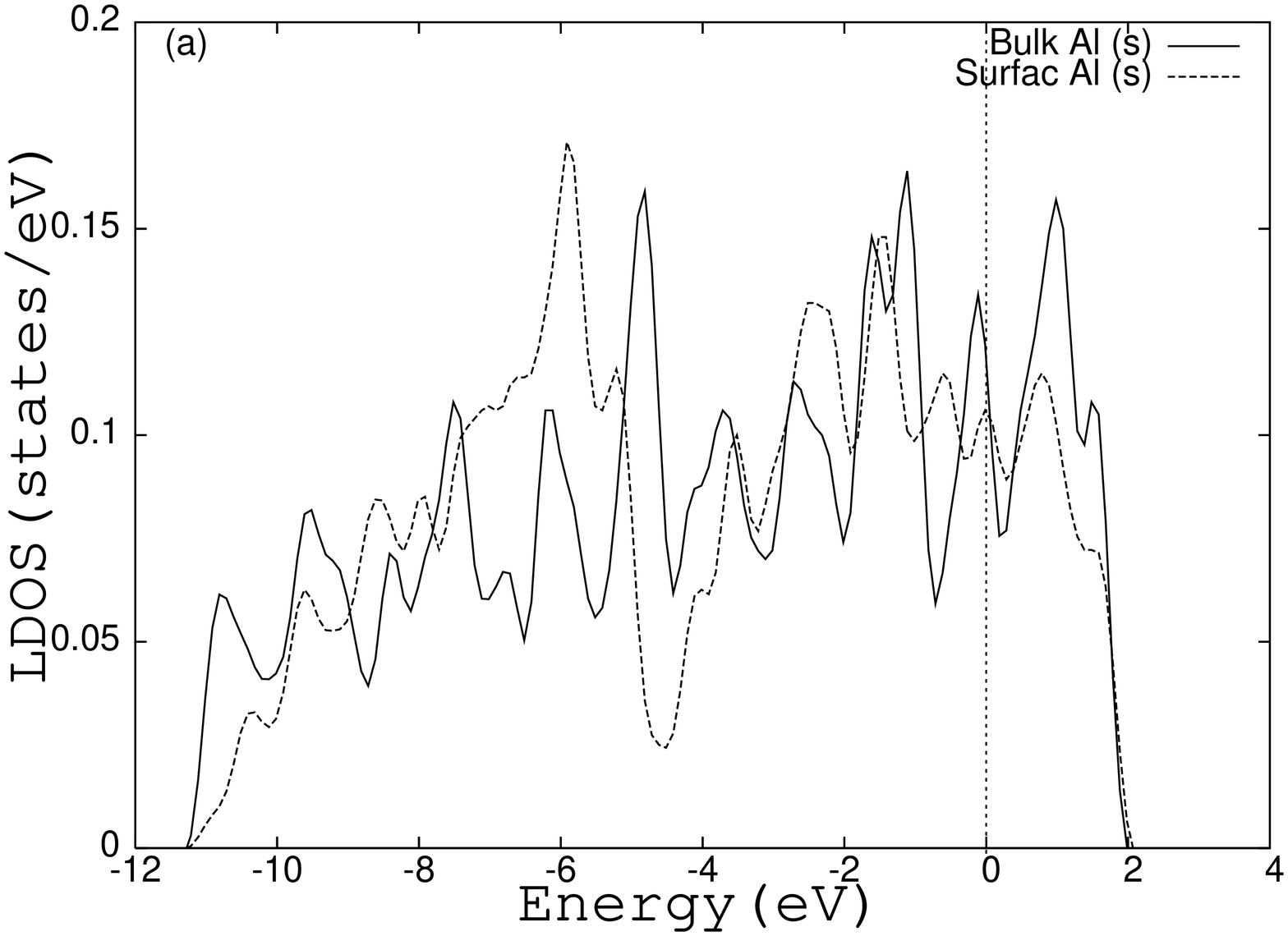}
\includegraphics[width=4cm,angle=-90]{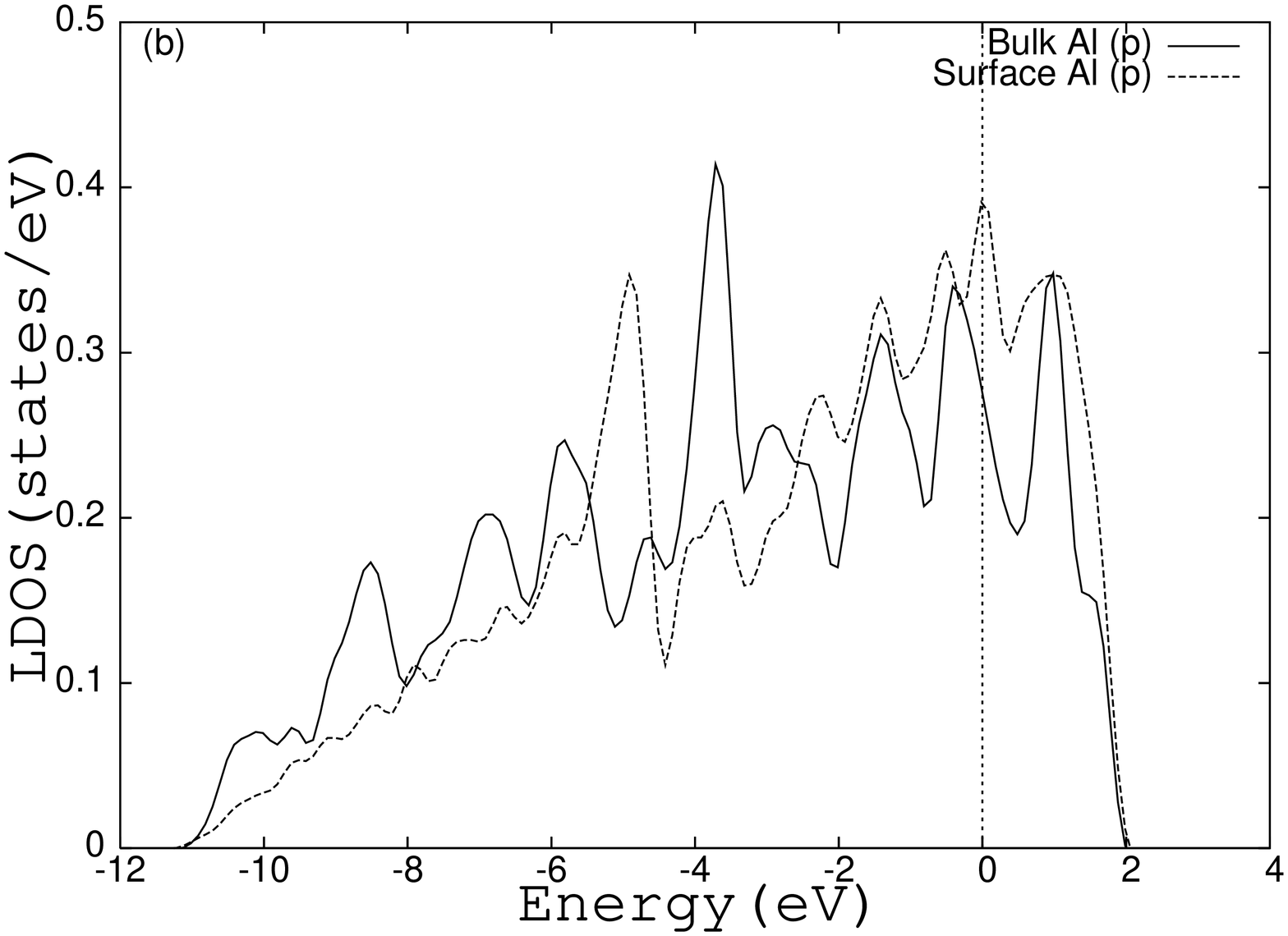}
\caption{\label{fig6}(a) and (b) represent LDOS for $s$ and $p$
orbitals, respectively, of Al atoms in the bulk (central layer of the slab) and surface layer of the Au/Al(001) system with $\Theta=0.25$~ML. }
\end{figure}

Analyzing the local density of states (LDOS) of Al atoms at the
central layer of the slab and Al atoms at surface shows some
shifts in energies and heights. The situation for coverage
$\Theta=0.25$~ML is shown in Figs.~\ref{fig6}~(a)-(b) for $s$ and $p$
orbitals, respectively. The Fermi energy of the Au/Al(001) system for
coverage 0.25~ML is obtained as 0.08~eV. As is seen in
Fig.~\ref{fig6}~(b), for energies near to Fermi energy, the density
of $p$ states shows an increase for the Al atoms in the surface
layer. On the other hand, in energies much lower than the Fermi energy,
the $p$-DOS decrease for the surface Al
atoms. However, the
$s$-DOS in Fig.~\ref{fig6}~(a) does not show much difference at the
Fermi surface, but some energy shifts at lower energies. Moreover, it is seen that the density of states (both $s$ and $p$) for the surface Al in the Au/Al(001) system are separated in to two groups with a gap around -4.5~eV. Conventionally, the lower-energy bunch is called as "bonding" states, and the higher one as "anti-bonding" states.

\begin{figure}
\includegraphics[width=4cm,angle=-90]{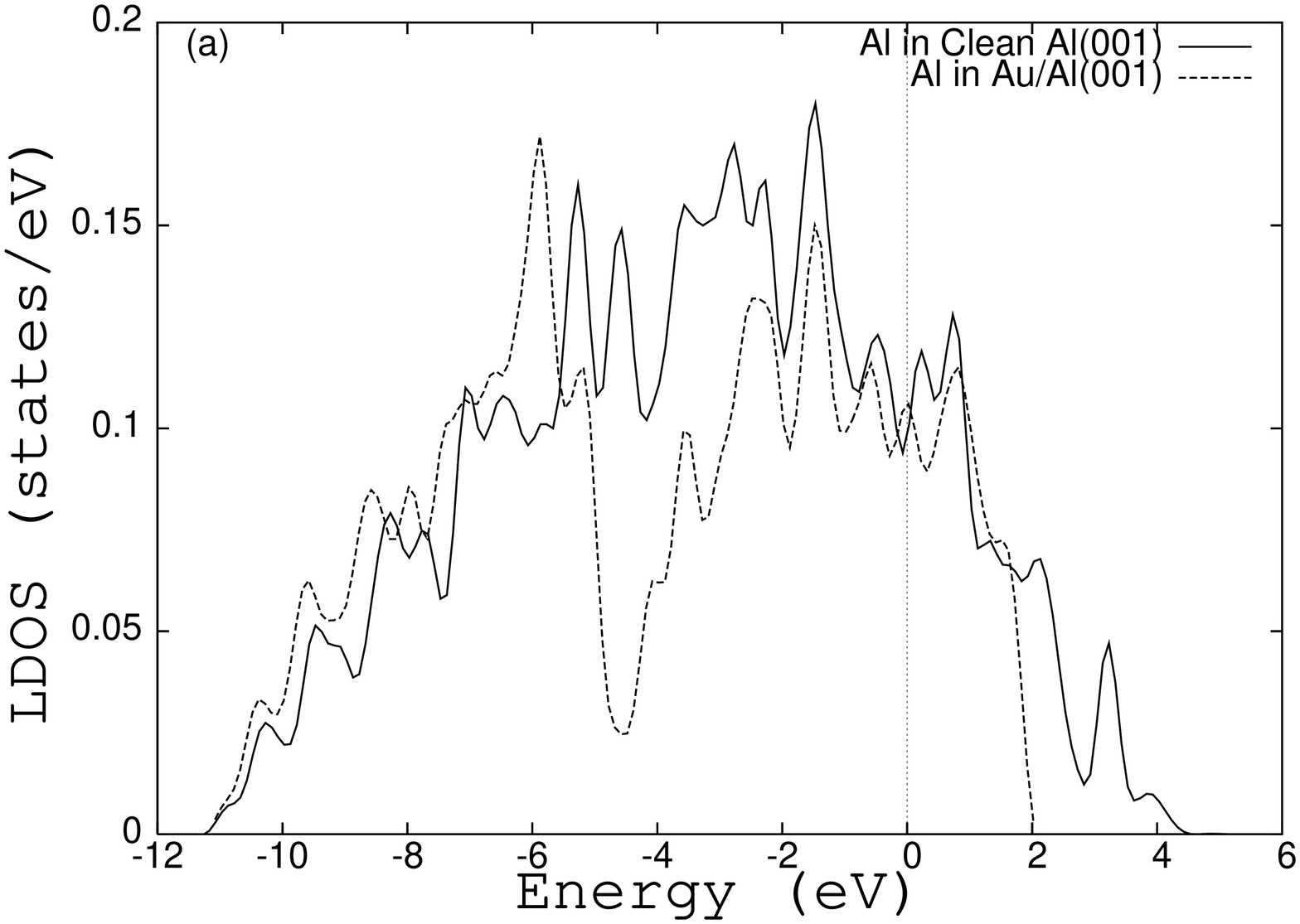}
\includegraphics[width=4cm,angle=-90]{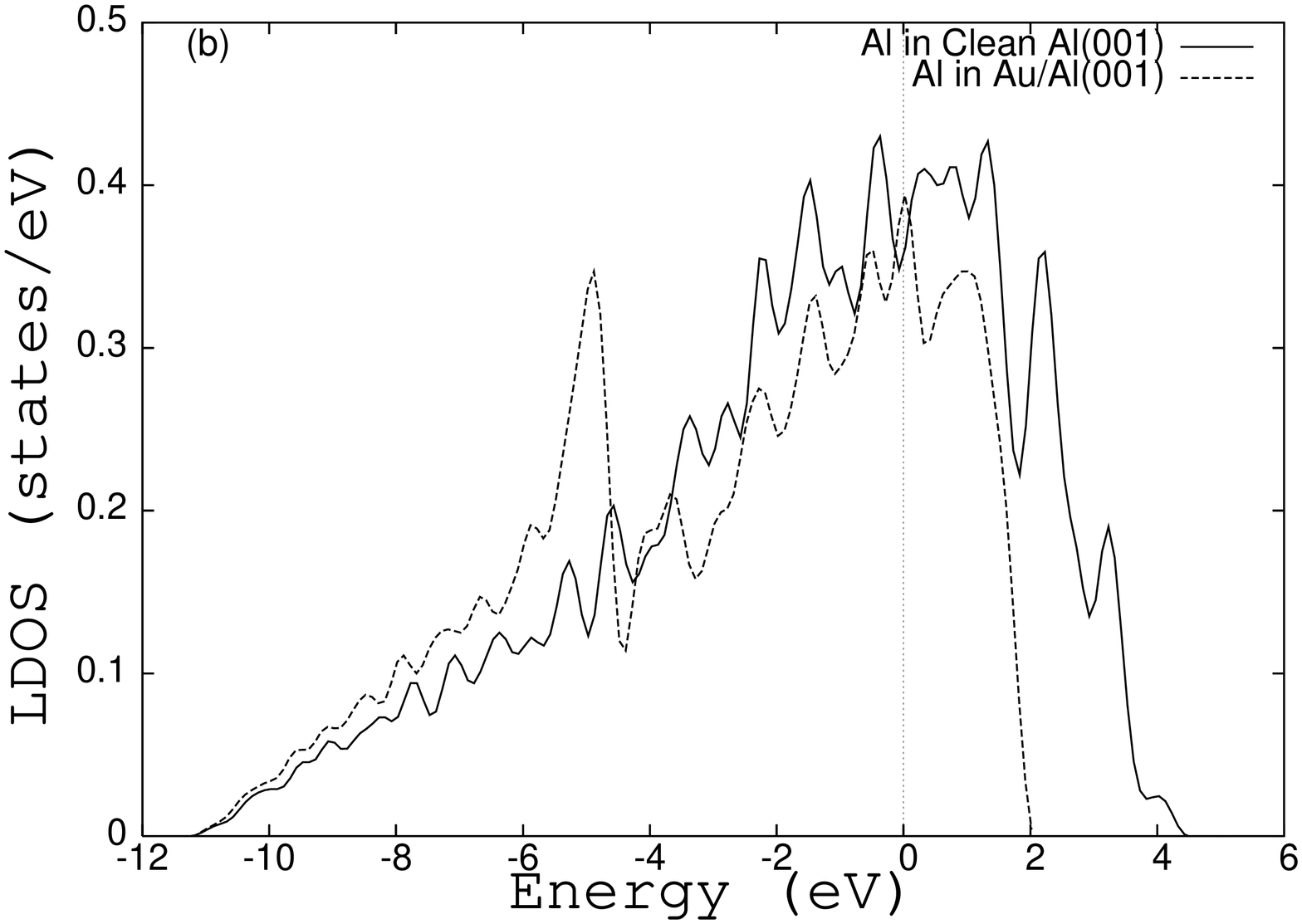}
\caption{\label{fig7}(a) and (b) represent LDOS for $s$ and $p$
orbitals, respectively, of the surface Al atoms in the clean Al(001) and Au/Al(001) surfaces for
$\Theta=0.25$~ML.  }
\end{figure}

In order to see the effects of Au adlayer on the surface Al atoms, we have compared, in Figs.~\ref{fig7}(a)-(b), the $s$- and $p$-DOS of the Al atoms in the clean Al(001) and the Au/Al(001) systems with coverage $\Theta$=0.25~ML. The plots show that the above-mentioned bunching appears because of the adsorbed Au atoms on the surface. The Fermi energy for the clean Al(001) system is -0.33~eV, and in the plots we have identified the Fermi energies as the zero of energy.  In Fig.~\ref{fig7}(a), the comparison of the $s$-DOS's shows that near the Fermi energy, there is a significant decrease for the Au/Al(001) system, while for lower energies ($\lesssim -5.5$~eV) there is a small increase. On the other hand, in Fig.~\ref{fig7}(b) we see that the $p$-DOS of the Au/Al(001) system has a small decrease near the Fermi surface, while it has an increase for lower energies ($\lesssim$-4.5~eV). Since, the $p$-DOS of the Au/Al(001) system has a peak at $\sim$~-5~eV, there will be a small increase of the $p$ charge on the Al atom in Au/Al(001) system.

\begin{figure}
\includegraphics[width=4cm,angle=-90]{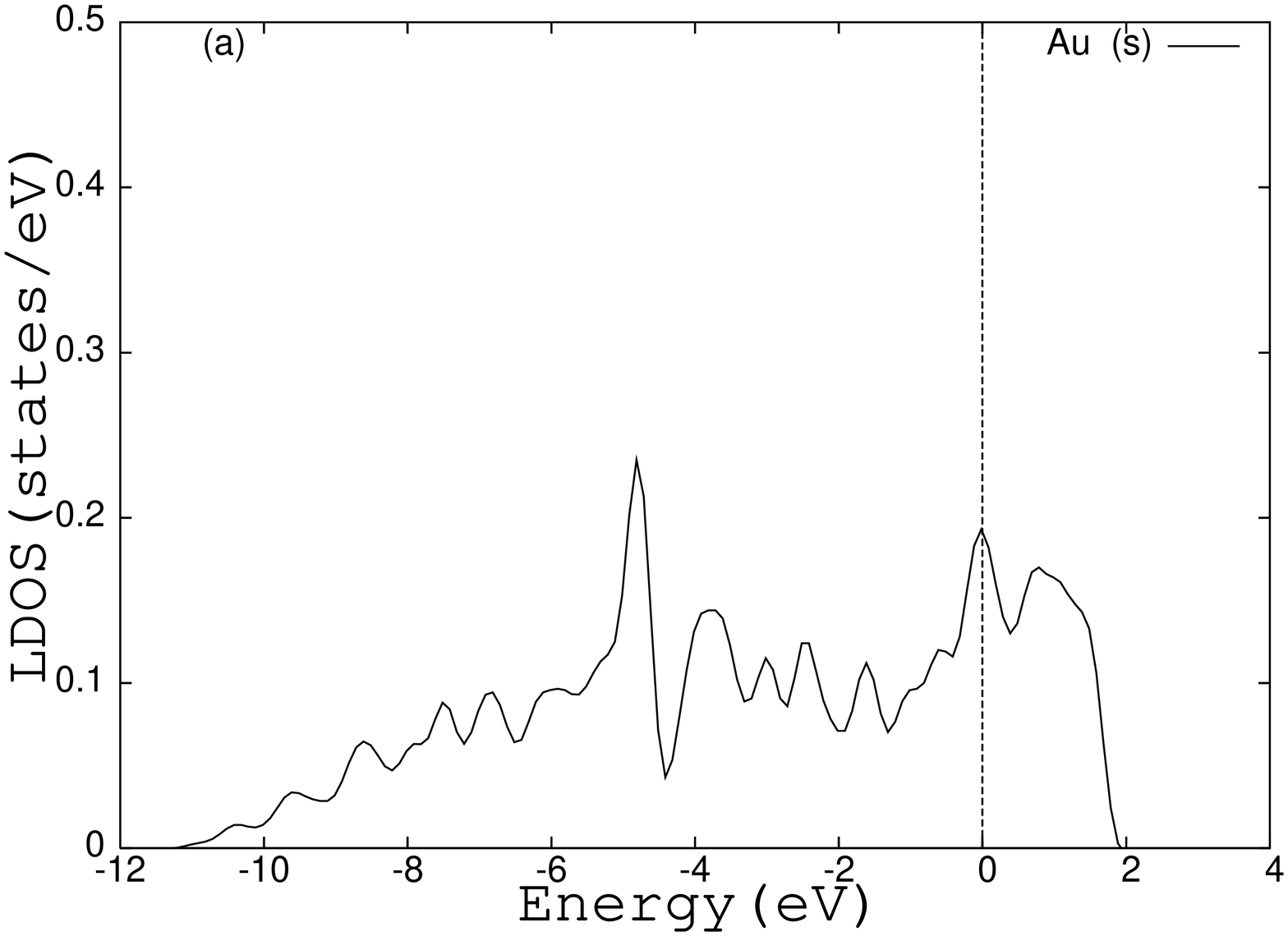}
\includegraphics[width=4cm,angle=-90]{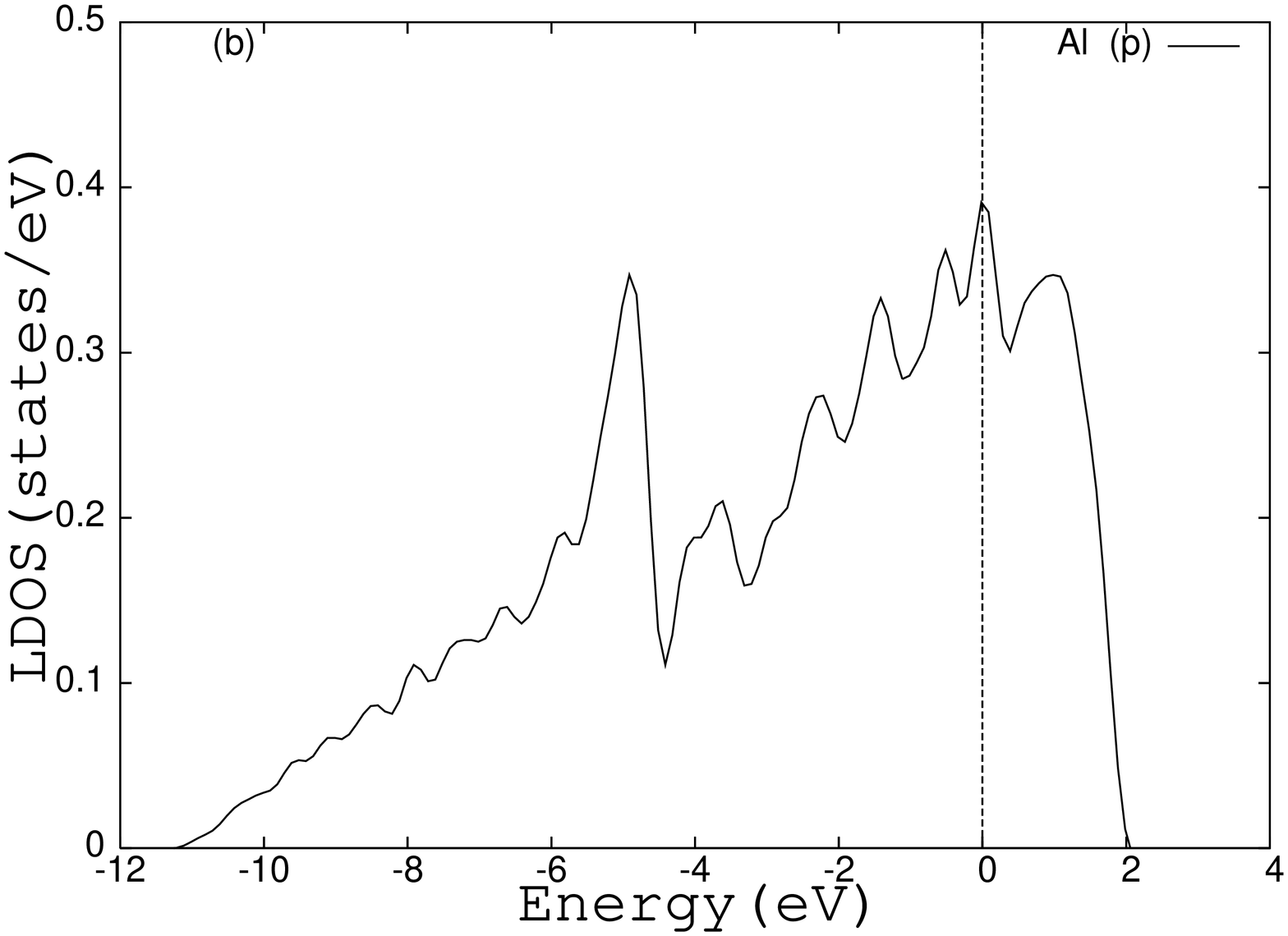}

\includegraphics[width=4cm,angle=-90]{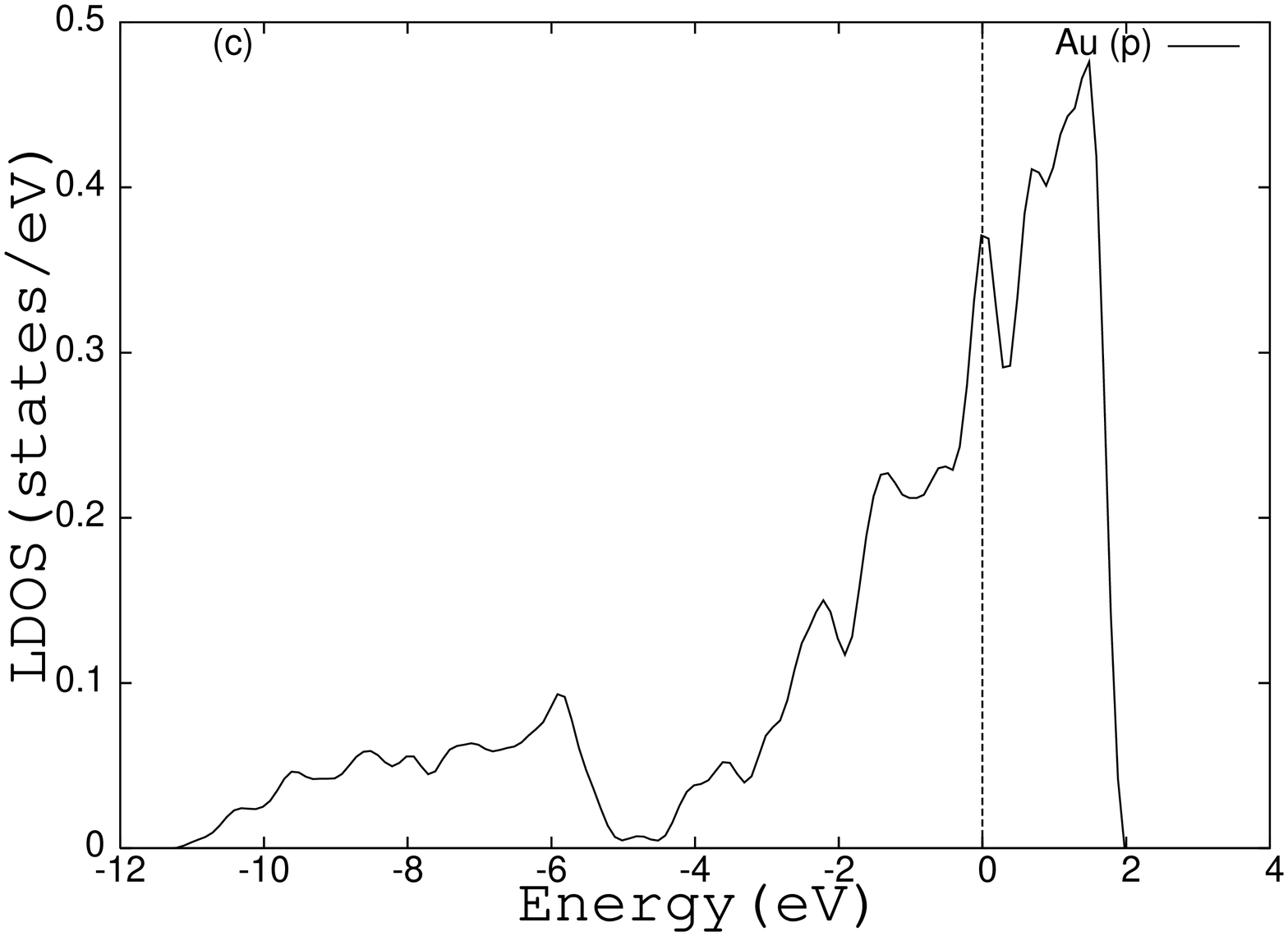}
\includegraphics[width=4cm,angle=-90]{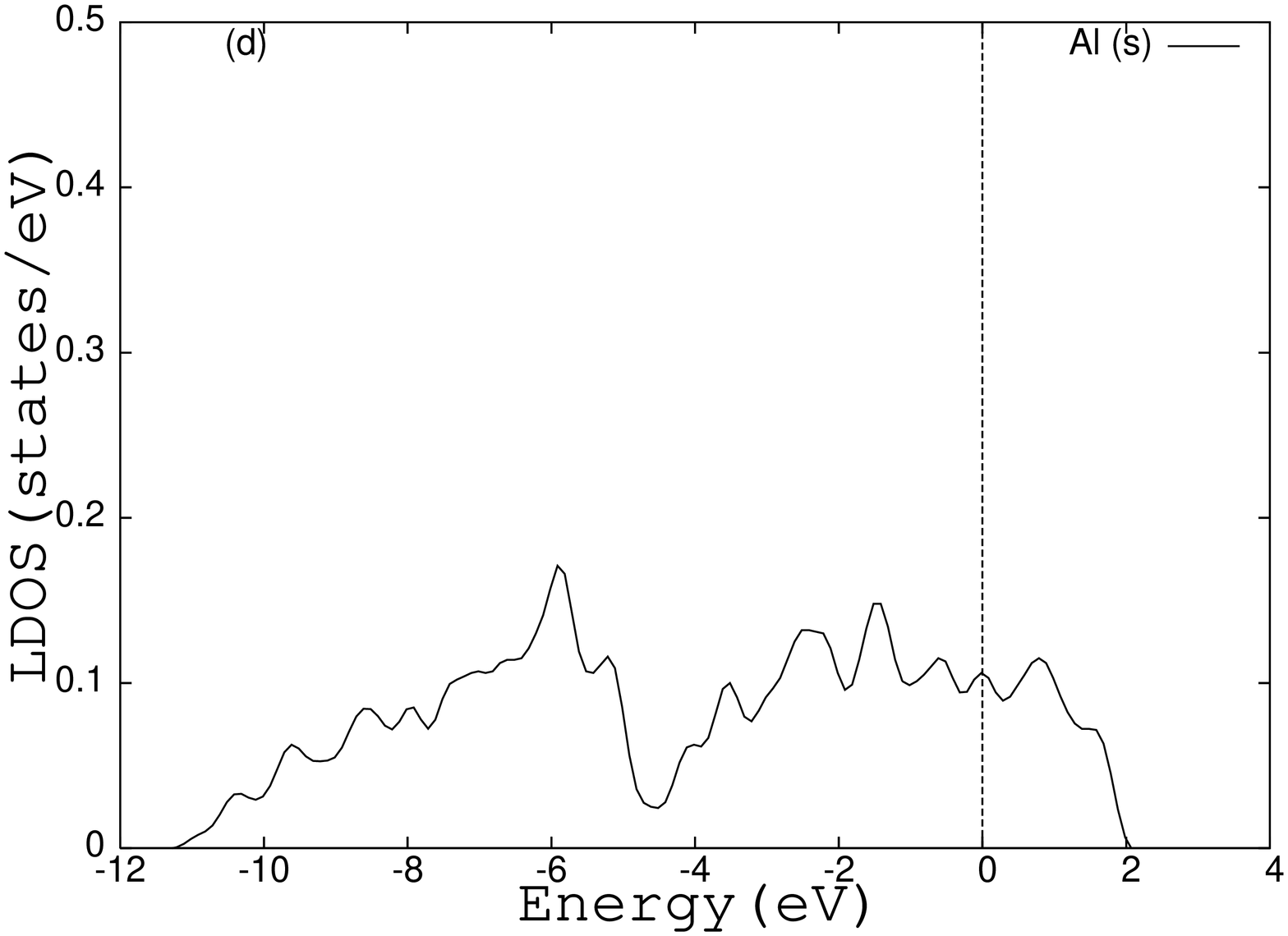}
\caption{\label{fig8}LDOS of the $s$ and $p$ orbitals for surface
Al and Au atoms for $\Theta=$0.25~ML. As is seen, the sharp peaks
of $s$(Au) and $p$(Al) indicate a strong hybridization between
the two orbitals. The Fermi energy is identified as the zero of
energy.}
\end{figure}

It is also instructive to compare the Au-DOS and the Al-DOS for
the nearest neighbor Al and Au atoms at surface. We have shown
this comparison for the case of $\Theta=0.25$~ML in
Fig.~\ref{fig8}(a)-(d).
Looking at Figs.~\ref{fig8}(a) and
\ref{fig8}(b) shows a significant overlap between $s$(Au) and $p$(Al)
on the one hand, and a significant overlap between $p$(Au) and $s$(Al), on the other
hand, as shown in Figs.~\ref{fig8}(c) and ~\ref{fig8}(d). Here we see that the bunching also happens in the Au-DOS's.
At the same time, the sharp peaks (at $\sim$~-5~eV) of
$s$(Au) and $p$(Al) and the sharp peaks (at $\sim$~-6~eV) of $p$(Au) and $s$(Al) indicate a strong hybridization between the corresponding two orbitals.

To analyze the bonding of Au in the Au/Al(001) system, we have further compared the $s$-DOS and $p$-DOS of the Au atom in the isolated Au adlayer with those corresponding to the Au atoms in the Au/Al(001) system for $\Theta$=0.25~ML. Our results for the isolated Au adlayer show that the $s$-DOS (with large peak) and the $p$-DOS (with a small peak) are centered at 0.16~eV and 0.06~eV with respect to the Fermi energy, and are partially filled. On the other hand, Figs.~\ref{fig8}(a) and \ref{fig8}(c) show that the electrons completely occupy the bonding levels and partially occupy the anti-bonding levels, which lie lower than the energy position of the peaks in the isolated adlayer system. This redistribution of the electrons to the lower energy levels leads to the bonding of the Au atoms on the Al(001) surface.

\begin{figure}
\includegraphics[width=5in]{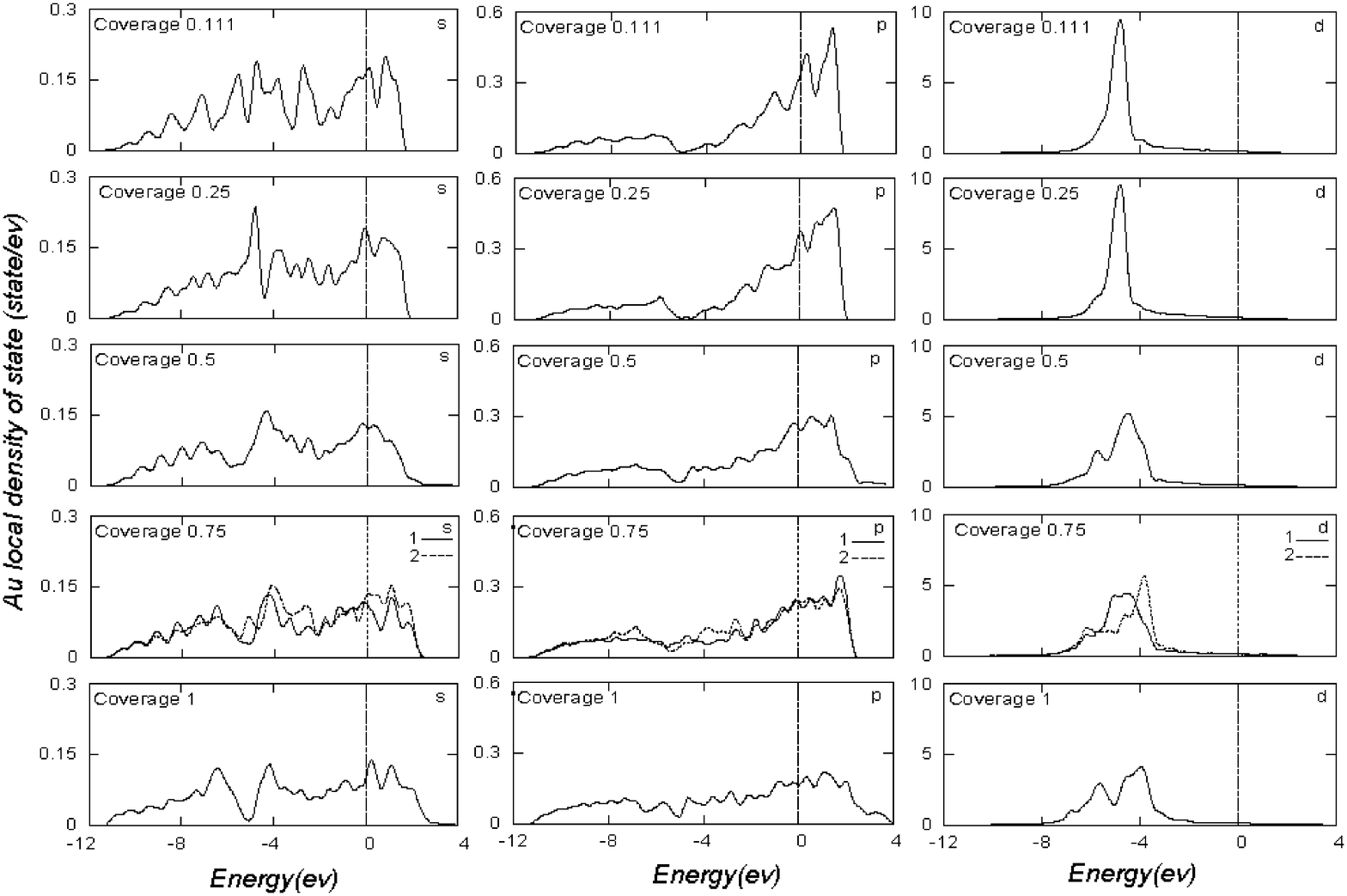}
\caption{\label{fig9}Evolution of the LDOS for the $s$, $p$, and
$d$ orbitals of Au adsorbed atom with different coverages. For
coverage $\Theta$=0.75~ML, because of two inequivalent Au atoms "1"
and "2", we have two curves in the plots which are slightly different in each
case. As is seen, with increasing $\Theta$, the density of states
undergo shifting and broadening. This broadening and shifting is
significant for the $d$(Au) orbital. }
\end{figure}

Finally, in Fig.~\ref{fig9} we have depicted the evolution of
LDOS for the adsorbed Au atoms with different coverages. In this
figure, the Fermi energy is identified as the zero of energy. For the system with
coverage $\Theta$=0.75~ML, because of two inequivalent Au atoms "1"
and "2", we have two curves in the plots which are slightly different in each
case. As is seen, with increasing $\Theta$, the density of states
undergo shifting and broadening. This broadening and shifting is
significant for the $d$(Au) orbital.

\section{\label{sec4}Conclusions}
In this work, we have studied the effects of Au adsorption on the
surface properties of the Al(001), by performing {\it ab initio}
calculations in the framework of density functional theory and
supercell methods.
Having calculated the adsorption energies, we have found the hollow sites on the surface as the most preferred adsorption sites, and
have considered the adsorption of Au atoms at the hollow sites
for different coverages from $\Theta=0.11$~ML to $\Theta=1.00$~ML.
We have shown
that the adsorbed Au atoms, because of higher electronegativity,
attracts the electronic charge of the substrate and cause the
induction of surface dipole moment. The induction of the surface dipole moment gives rise to the work-function change.
Using the Helmholtz relation, we have obtained a sharp dip in the induced dipole moment at $\Theta$=0.25~ML. We have shown that using the self-consistent electronic density and taking into account the ionic charge redistributions, one can directly calculate the induced surface dipole moment. Therefore, the sharp dip is not only due to electronic charge redistributions but, one should take into account the ionic redistributions as well.
Moreover, we have shown that, at zero temperature, the Au atoms do not diffuse into the Al(001) substrate, in agreement with the experimental observations.
To gain an insight on the nature of bondings, we have also studied the local density of
states. Finally, for sufficiently high Au coverages, one has to consider the clustering effects. The clustering leads to the formation of Au nano-clusters on the Al(001) surface which is important in catalytic applications. Work in this direction is in progress.

\section*{Acknowledgments}
MP would like to thank Professor J.~W.~F.~Egelhoff for providing a copy of his published work on Au/Al and useful discussion. This work is part of research program in NSTRI, Atomic Energy Organization of Iran.

\end{document}